\documentclass[a4paper]{amsart}

\usepackage{graphicx}
\usepackage{a4wide,verbatim}
\usepackage[dvipsnames]{xcolor}
\usepackage{epstopdf, epsfig}
\usepackage{amsaddr}
\usepackage{amsmath,amsfonts,amssymb,amsgen,amsbsy,dsfont}
\usepackage{floatrow}
\usepackage{subfig}
\usepackage{caption}
\usepackage{hyperref}

\newcommand{\depth}{d}                              
\newcommand{\sur}[1]{{#1}_\mathrm{s}}                    
\renewcommand{\bot}[1]{{#1}_\mathrm{b}}                  
\newcommand{\eqdef}{\stackrel{\text{\tiny{def}}}{=}} 

\newcommand{\ud}{\mathrm{d}}

\newcommand{\ue}{\mathrm{e}}
\newcommand{\ui}{\mathrm{i}}

\newcommand{\half}{{\textstyle{\frac{1}{2}}}}

\renewcommand{\Re}{\operatorname{Re}}
\renewcommand{\Im}{\operatorname{Im}}

\newlength{\intwidth}

\begin{document}

\title[Recovery of water-waves with arbitrary surface-pressure]{Steady water-waves with arbitrary surface-pressure: Their recovery from bottom-pressure measurements}

\author[D. Clamond]{Didier Clamond}
\address{Universit\'e C\^ote d'Azur, CNRS UMR 7351,  Laboratoire J. A. Dieudonn\'e, 
Parc Valrose, 06108 Nice cedex 2, France}
\email{didier.clamond@univ-cotedazur.fr}

\author[J. Labarbe]{Joris Labarbe}
\address{Universit\'e C\^ote d'Azur, CNRS UMR 7351,  Laboratoire J. A. Dieudonn\'e, 
Parc Valrose, 06108 Nice cedex 2, France}
\email{jlabarbe@unice.fr}

\subjclass[]{}

\begin{abstract}
Equations relating the pressure at a horizontal seabed, the free-surface 
profile and the surface-pressure are derived for two-dimensional irrotational 
steady water waves with arbitrary pressure at the free surface. Special cases 
include gravity, capillary, flexural and wind waves. Without approximations, 
we show that the free-surface recovery from the bottom-pressure requires the 
resolution of only one first-order ordinary differential equation independent 
of the surface-pressure, thus providing a new general recovery method valid for a 
broad class of water waves. 
Another equation provides an explicit expression for the surface-pressure as 
a function of the bottom-pressure and of the free-surface. Thus, if unknown, 
the surface-pressure can be also recovered if one extra measurement is available. 
This new recovery procedure is illustrated analytically for the linear approximation 
of a flexural-capillary-gravity wave, and numerically for fully nonlinear 
capillary-gravity waves.
\end{abstract}

\maketitle

\section{Introduction}

In this paper, we present equations relating the surface-wave profile, the 
surface-pressure and the bottom-pressure. This study includes any surface-pressure 
describing various physical effects, such as capillarity, flexural elasticity, 
wind stress, etc. 

Methods for recovering pure gravity (i.e., with constant pressure at the free 
surface) irrotational waves from bottom pressure gauges have long been proposed.  
These methods either solve the problem exactly or under various simplifications; 
see \cite{Constantin2012b}, \cite{OliverasEtAl2012}, \cite{ClamondConstantin2013}, \cite{VasanEtAl2017} and the references 
therein for reviews and details. Recently, \cite{ClamondEtAl2023} showed that 
an exact recovery is also possible in presence of constant vorticity. However, 
to the present authors knowledge, the recovery of capillary, flexural and wind waves 
(among many other situations of physical interest) has never been attempted. 
These phenomena involve different non-constant surface-pressures 
that can be very complicated (especially for capillary and 
flexural waves), and the surface-pressure is generally a function 
of the free surface profile that is unknown {\em a priori}. Hence, 
compared to the case with constant surface-pressure (i.e., pure 
gravity waves) treated in the references cited above, considering  
varying surface-pressure is a major additional complication, requiring 
a new method of resolution for the wave recovery problem.

In this short paper, we describe a new general recovery method valid 
for any surface-pressure. This is possible because the free-surface 
recovery from the bottom-pressure requires 
the resolution of only one first-order ordinary differential equation independent 
of the surface-pressure. Once known, the surface-profile and the bottom-pressure 
yield an explicit relation for the surface-pressure. Thus, the surface-profile 
and the surface-pressure are both determined from the bottom-pressure, but modulo 
an unknown scalar parameter (e.g., the Bernoulli constant), so one extra 
relation is required to close the problem. This can be obtained either by an 
extra measurement or by the knowledge of the physical effects at the free-surface 
(i.e., knowing an equation the surface-pressure must satisfy). 

The paper is organised as follow.
Section \ref{secmathdef} is devoted to the physical assumptions and the resulting 
fundamental equations. Equations for the free-surface and the surface-pressure 
recovery from the bottom-pressure are derived in section \ref{secrecov}. The 
recovery procedure is illustrated analytically and numerically in sections 
\ref{appapp} and \ref{secex2}, respectively. Finally, the section \ref{secconclu} outlines some conclusions 
and perspectives.
  
\section{Preliminaries}\label{secmathdef}

In the frame of reference moving with a traveling wave of permanent shape, the 
flow beneath the wave is a steady two-dimensional irrotational motion of an 
inviscid fluid. Note that the wave phase velocity $c$ is a non-zero constant in 
any other Galilean frame of reference. 
Let be $(x,y)$ a Cartesian coordinate system moving with the wave, $x$ being the 
horizontal coordinate and $y$ being the upward vertical one, and let be  
$\left(u(x,y),\,v(x,y)\right)$ the velocity field in this moving frame of reference. 
We denote by $y=-\depth$, $y=\eta(x)$ and $y=0$ the equations at the bottom, the 
free surface and the mean water level, respectively. 
The latter equation expresses that $\left<\eta\right>=0$ for a smooth 
$(2\pi/k)$-periodic wave profile $\eta$, where $\left<\cdot\right>$ is the 
Eulerian average operator over one period, i.e.
\begin{equation} \label{defmean}
\left<\eta\right> \eqdef \frac{k}{2\pi} \int_{-{\pi/k}}^{{\pi/k}} \eta(x) \ud\/x = 0.
\end{equation}
For solitary and more general aperiodic waves, the same averaging operator 
applies taking the limit $k\to0^+$.  

The flow is governed by the balance between the restoring gravity force, the 
inertia of the system and a surface-pressure. With constant density $\rho>0$ and 
acceleration due to gravity $g>0$ directed downward, the kinematic and dynamic 
equations  are, for $(x,y)\in\mathds{R}\times[-\depth,\eta(x)]$ 
\cite{WehausenLaitone1960},
\refstepcounter{equation}
\[
u_x + v_y = 0, \quad
v_x - u_y = 0, \quad
uu_x + vu_y = -P_x/\rho, \quad
uv_x + vv_y = -P_y/\rho - g,
\eqno{(\theequation{\mathit{a},\mathit{b},\mathit{c},\mathit{d}})}\label{eqeul}
\]
where $P(x,y)$ denotes the hydrodynamical pressure. 

The flat bottom and steady free surface being impermeable, we have
\refstepcounter{equation}
\[
\bot{v} = 0, \qquad
\sur{v} = \sur{u} \eta_x,
\eqno{(\theequation{\mathit{a},\mathit{b}})}\label{impcond}
\]
with $\eta_x\eqdef\ud\eta/\ud x$ and where subscripts `$\text{b}$' and `$\text{s}$' 
denote, respectively, restrictions at the bottom and at the free surface, e.g. 
$\bot{u}(x)=u(x,-\depth)$, $\sur{v}(x)=v(x,\eta(x))$. The pressure at the free 
surface is  
\begin{equation}
\label{d}
\sur{P} = P_\text{atm} + \rho \sur{p} \qquad \hbox{at} \quad y = \eta(x),
\end{equation}
where $P_\text{atm}$ is a constant atmospheric pressure and $\sur{p}$ is a 
varying pressure (divided by the density). For instance, one can consider a 
prescribed surface pressure such as a Gaussian distribution of magnitude $p_0$ 
and variance $\lambda$ \cite{WadeEtAl2014}
\begin{equation} 
\label{pregau}
\sur{p} = p_0 \exp\!\left[-x^2/(2\lambda)\right],
\end{equation} 
or capillary and flexural effects such that \cite{LAmb1932,Toland2008} 
\begin{equation}
\label{defpscapflex}
\sur{p} = -\frac{\ud}{\ud\/x} \left\{ \frac{\tau\eta_x}{\left(1+\eta_x^{2}
\right)^{1/2}} - \frac{D\eta_{xxx}}{\left(1+\eta_x^{2}\right)^{5/2}} +
\frac{5D\eta_{x}\eta_{xx}^{2}}{2\left(1+\eta_x^{2}\right)^{7/2}} \right\}, 
\end{equation}
$\tau$ being a surface tension coefficient and $D$ a rigidity parameter 
(both divided by the fluid density). Other phenomena can of course be considered, 
as well as their combination. Without loss of generality, we take 
$\left<\sur{p}\right>=0$ since $\left<\sur{p}\right>$ can be absorbed into the 
definition of $P_\text{atm}$. It is thus convenient to introduce the normalised 
relative pressure
\begin{equation}\label{pd}
p(x,y) \eqdef \left[ P(x,y)-P_\text{atm} \right]/\rho, \qquad 
(x,y)\in\mathds{R}\times[-\depth,\eta(x)].
\end{equation}

The flow being irrotational, the dynamical (Euler) equations (\ref{eqeul}{\it 
c-d}) can be integrated into a Bernoulli equation
\begin{equation}\label{bernbase}
2 (p + gy) + u^2 + v^2 = B,
\qquad (x,y)\in\mathds{R}\times[-\depth,\eta(x)] ,
\end{equation}
where $B$ is a Bernoulli constant. From equations \eqref{defmean}--\eqref{d} 
and \eqref{bernbase}, one gets \cite{ClamondConstantin2013,ClamondEtAl2023}
\begin{equation}
\label{mt}
B = \left<\sur{u}^{2} + \sur{v}^{2} \right> = \left< \bot{u}^{2} \right>,
\end{equation}
yielding the, here important, relation 
\begin{equation}\label{meanpb}
\left<\bot{p}\right> = g\depth.
\end{equation}

Finally, equations (\ref{eqeul}{\it a-b}) imply that the complex velocity 
$w\eqdef u-\ui v$ is a holomorphic function of the complex coordinate $z\eqdef 
x+\ui y$, an interesting feature exploited below.

\section{Equations for the free-surface and surface-pressure recoveries}\label{secrecov}

For free-surface and surface-pressure recoveries, we present here a simple  
derivation of equations generalising those of \cite{Clamond2013} and 
\cite{ClamondConstantin2013}. 

\subsection{General equations} 

The function $w^2$ being holomorphic, its real and imaginary parts satisfy 
the Cauchy--Riemann relations 
\refstepcounter{equation}
\[
\partial_y\!\left(u^2-v^2\right) - \partial_x\!\left(2uv\right) = 0, \qquad
\partial_x\!\left(u^2-v^2\right) + \partial_y\!\left(2uv\right) = 0.
\eqno{(\theequation{\mathit{a},\mathit{b}})}\label{CRw2B}
\]
Integrating over the water column and using the boundary 
conditions, these relations yield after some elementary algebra 
\refstepcounter{equation}
\[
\bot{p} - \sur{p} - gh = \frac{\ud}{\ud x} \int_{-\depth}^\eta
uv\/\ud y, \qquad
\left( \sur{p} + g\eta \right) \frac{\ud\eta}{\ud x} = \frac{\ud}{\ud x}
\int_{-\depth}^\eta \frac{u^2-v^2+B}{2}\/ \ud y. 
\eqno{(\theequation{\mathit{a},\mathit{b}})}\label{intCRw2B}
\]

Taylor expansions around $y=-\depth$ can be written 
\cite{Clamond2022,Fenton1972,Lagrange1781} 
\begin{align}
u^2 - v^2 &= \cos\!\left[ (y+\depth) \partial_x \right] \bot{u}^{2}
 = -2 \cos\!\left[ (y+\depth) \partial_x \right] (\bot{p} - g\depth), \\
2uv &= -\sin\!\left[ (y+\depth) \partial_x \right] \bot{u}^{2} 
= 2 \sin\!\left[ (y+\depth) \partial_x \right] (\bot{p}-g\depth)
\end{align}
or in complex form
\begin{equation}
w(z)^2 = \exp\!\left[ \ui (y+\depth) \partial_x \right] \bot{u}(x)^2 
= \bot{u}(z+\ui\depth)^2 = B + 2g\depth - 2\bot{p}(z+\ui\depth).
\end{equation}
(For any real function $F(x)$ continuable in the complex plane, $F(x+\ui h)=
\exp\!\left[ \ui h \partial_x \right]F(x)$ is the Taylor expansion around 
$h=0$.) Hence, with $h\eqdef\depth+\eta$, we have
\begin{subequations}\label{intCRpb}
\begin{align}
\int_{-\depth}^\eta u\/v\/ \ud\/y &= \left[ 1 - \cos\!\left( h
\partial_x \right) \right] \partial_x^{-1} \left(\bot{p} - g\depth\right), \\
\int_{-\depth}^\eta \frac{u^2-v^2+B}{2}\/ \ud\/y &= -\sin\!\left( h
\partial_x \right) \partial_x^{-1} (\bot{p} - g\depth),
\end{align}
\end{subequations}
so equations \eqref{intCRw2B} yield
\begin{align}
\sur{p} + g\eta &= \partial_x \cos\!\left( h \partial_x \right)
\partial_x^{-1} \left(\bot{p} - g\depth \right) = \left[ 
\cos\!\left( h \partial_x \right) - \eta_x \sin\!\left( h \partial_x
\right) \right] (\bot{p} - g\depth), \label{CC1} \\
(B-\sur{p}-g\eta) \eta_x &= \partial_x \sin\!\left( h \partial_x \right)
\partial_x^{-1} \left(\bot{p} - g\depth \right) = \left[ 
\sin\!\left( h \partial_x \right) + \eta_x \cos\!\left( h \partial_x
\right) \right] (\bot{p} - g\depth). \label{CC2}
\end{align}

After one integration, equation \eqref{CC2} becomes
\begin{align}
B\eta - \half g \eta^2 - \partial_x^{-1} \left( \sur{p} \eta_x \right) = 
\sin\!\left( h \partial_x \right) \partial_x^{-1} (\bot{p}-g\depth) . \label{CC3}
\end{align}
With the special surface pressure \eqref{pregau} the term $\partial_x^{-1}\sur{p}
\eta_x$ cannot be obtained in closed form, but with \eqref{defpscapflex} we have
\begin{equation}\label{Ipex}
\partial_x^{-1} \left( \sur{p} \eta_x \right) = \frac{\tau}
{\left(1+\eta_x^{2}\right)^{1/2}}\,-\,\tau
+ \frac{D \eta_x \eta_{xxx} - 3D \eta_{xx}^{2}}{\left(1+\eta_x^{2}\right)^{5/2}}
+ \frac{5D \eta_{xx}^{2}}{2\left(1+\eta_x^{2}\right)^{7/2}} + \text{constant},
\end{equation}
where the integration constant must be determined by the mean level condition 
\eqref{defmean}, i.e., imposing
\begin{align}
\left< \half g \eta^2 + \partial_x^{-1} \left( \sur{p} \eta_x \right) +
\sin\!\left( h \partial_x \right) \partial_x^{-1} (\bot{p} - g\depth)
\right> = 0 . \label{CC3ave}
\end{align}
Note that the value of the integration constant in $\partial_x^{-1} 
(\bot{p}-g\depth)$ does not matter here because this constant vanishes 
after application of the pseudo-differential operator $\sin\!\left( h 
\partial_x \right)$.    

Equations \eqref{CC1}, \eqref{CC2} and \eqref{CC3} are generalisations 
for $\sur{p}\neq0$ of the relations derived by 
\cite[eq. 3.5--3.6]{ClamondConstantin2013} and by 
\cite[eq. 4.4]{Clamond2013} when $\sur{p}=0$. (This is obvious introducing 
the holomorphic function $\mathfrak{P}(z) \eqdef \bot{p}(z+\ui\depth)$ and 
$\mathfrak{Q}(z)\eqdef \int\left[\mathfrak{P}(z)-g\depth\right] \ud z$.) 

\subsection{Generic equation for the free-surface recovery}

When $\sur{p}=0$ (pure gravity waves), $\eta$ can be obtained from $\bot{p}$ 
solving the ordinary differential equation \eqref{CC2} \cite{ClamondConstantin2013} 
or, more easily, solving the algebraic equation \eqref{CC3} \cite{Clamond2013}. 
When $\sur{p}\neq0$ is a function of $x$ and/or $\eta$, such as \eqref{pregau} 
and \eqref{defpscapflex}, in general \eqref{CC3} is a complicated highly-nonlinear 
high-order integro-differential equation for $\eta$ due to the term 
$\partial_x^{-1}\left( \sur{p} \eta_x \right)$ (see relation \eqref{Ipex} for 
an example of practical interest).  
This is not a problem for recovering the free surface $\eta$ 
from the bottom-pressure $\bot{p}$ because the surface pressure $\sur{p}$ can be 
eliminated between \eqref{CC1} and \eqref{CC2}, yielding
\begin{equation}\label{CC4}
B \eta_x = \left[ \left(1-\eta_x^{2}\right) \sin\!\left(h \partial_x
\right) + 2 \eta_x \cos\!\left( h \partial_x \right) \right]
(\bot{p} - g\depth), 
\end{equation}
or in complex form --- introducing $\widetilde{\mathfrak{P}}(z)\eqdef
\bot{p}(z+\ui\depth)-g\/\depth$ ---
\begin{equation}\label{CC5}
B \eta_x = \left( 1-\eta_x^{2} \right) \Im\!\big\{ \sur{\widetilde{\mathfrak{P}}}
\big\} + 2\eta_x \Re\!\big\{ \sur{\widetilde{\mathfrak{P}}} \big\} ,
\end{equation}
that is a (nonlinear) first-order ordinary differential equation for $\eta$.
Equation \eqref{CC5} being algebraically quadratic for $\eta_x$, it can be 
solved explicitly for $\eta_x$; thus one gets 
\begin{equation}\label{recoveq}
\Re\!\big\{ \sur{\widetilde{\mathfrak{P}}} \big\} - \eta_x \Im\!
\big\{ \sur{\widetilde{\mathfrak{P}}} \big\}
= \half B \pm \half \vert B - 2 \sur{\widetilde{\mathfrak{P}}} \vert.
\end{equation}
Since the free surface is flat if the bottom pressure is constant (and because 
$B>0$), the minus sign must be chosen. Moreover, the condition \eqref{mt} 
rewritten in terms of $\widetilde{\mathfrak{P}}$ yielding 
$B = \langle \vert B - 2 \sur{\widetilde{\mathfrak{P}}} \vert \rangle$,
the average of the right-hand side of \eqref{recoveq} is zero, so is the 
left-hand side. 

Equation \eqref{recoveq} is {\em a priori} not suitable if 
$\eta$ is (nearly) not differentiable (limiting waves). It 
is thus more efficient to solve its antiderivative 
\begin{equation}\label{recoveqin}
\Re\!\left\{ \sur{\mathfrak{Q}} \right\} - K = \half\,\partial_x^{-1}
\left(B-\left|B - 2\/\sur{\widetilde{\mathfrak{P}}} \right|\right),
\end{equation}
where $K$ is an integration constant and where $\mathfrak{Q}(z)\eqdef
\bot{q}(z+\ui\depth)$, with $\bot{q}(x)\eqdef\partial_x^{-1}\/(\/\bot{p}
-g\/\depth\/)$. Assuming $\langle\bot{q}\rangle\eqdef0$ (without loss of 
generality), it yields 
$\partial_x\/\Re\!\left\{\sur{\mathfrak{Q}}\right\} = \Re\!\big\{ 
\sur{\widetilde{\mathfrak{P}}} \big\}
- \eta_x \Im\!\big\{ \sur{\widetilde{\mathfrak{P}}} \big\}$ and 
$\left<(1+\ui\eta_x)\sur{\mathfrak{Q}}\right>=0$. 
The right-hand side of  \eqref{recoveqin} being the antiderivative of a zero-average 
quantity, we conveniently choose $\langle \partial_x^{-1} \big(B - \vert B - 2
\sur{\widetilde{\mathfrak{P}}} \vert \big) \rangle \eqdef0$, hence 
$K = \langle \Re\!\big\{ \sur{\mathfrak{Q}} \big\} \rangle$. 
Thus, a numerical resolution of \eqref{recoveqin} does not require the computation 
of $\eta_x$, that is an interesting feature for steep waves. 

\subsection{Recovery of the surface-pressure}

The free-surface $\eta$ being obtained after the resolution of \eqref{recoveq} 
or \eqref{recoveqin}, the surface-pressure $\sur{p}$ is obtained explicitly at 
once from \eqref{CC1}
\begin{equation}\label{psrecov}
\sur{p} = \partial_x \Re\!\left\{\sur{\mathfrak{Q}}\right\} - g\eta
= \Re\!\big\{ \sur{\widetilde{\mathfrak{P}}} \big\} - \eta_x \Im\!\big\{ 
\sur{\widetilde{\mathfrak{P}}} \big\} - g \eta.
\end{equation}
Thus, as $\eta$, $\sur{p}$ is known modulo the Bernoulli constant $B$. 
Relation \eqref{meanpb} holds as a definition of the mean depth $\depth$, 
leaving us with only one scalar quantity to be determined (i.e., $B$). 

\subsection{Closure relation}\label{secclosure}  

In order to fully recover both the free-surface and the surface-pressure, 
knowing the bottom pressure is not sufficient and one extra information 
is needed. We consider here two possibilities of practical interest.

A first possibility is when we have access to one independent extra measurement, 
for instance the mean velocity at the bottom (or elsewhere), the mean pressure 
somewhere above the seabed, the phase speed, the wave height, etc. 
In that case, the Bernoulli constant $B$ is chosen such that the recovered 
wave matches this measurement. Thus, the free-surface and the surface-pressure 
can be both fully recovered. 

If no extra measurements are available (only the bottom pressure is known), 
the free-surface can nevertheless be fully recovered with the knowledge of 
the physical nature of the surface-pressure, for instance given by \eqref{pregau} 
or \eqref{defpscapflex} (among many other possibilities). The missing parameter 
can then be obtained minimising an error (quadratic or minimax, for example)
between the recovered surface-pressure 
${\sur{p}}_r$ obtained from \eqref{psrecov} and the theoretical surface-pressure 
${\sur{p}}_t$ given, say, by \eqref{defpscapflex}. 



\subsection{Remarks} 

The fact $\sur{p}$ can be eliminated is not surprising. Indeed, $\bot{p}$ too 
can be eliminated between \eqref{CC1} and \eqref{CC2}, yielding the equation 
\begin{equation}\label{CCeta}
\sur{p} + g\eta = \partial_x \cos\!\left( h\partial_x \right)
\sin\!\left( h\partial_x \right)^{-1}
\left[ B\eta - \half g\eta^2 - \partial_x^{-1}\left(\sur{p}\eta_x\right) \right], 
\end{equation}
or, after inversion of the pseudo-differential operator,
\begin{equation}\label{CCetabis}
B\eta - \half g\eta^2 - \partial_x^{-1}\left(\sur{p}\eta_x\right) = 
\sin\!\left(h\partial_x\right)\cos\!\left( h\partial_x \right)^{-1} 
\partial_x^{-1} \left( \sur{p} + g\eta \right).
\end{equation}
Relation \eqref{CCetabis} with $\sur{p}=0$ is an Eulerian counterpart of the 
\cite{Babenko1987} equation \cite{Clamond2018}. A more involved Eulerian 
equation, somehow similar to \eqref{CCeta} with $\sur{p}=0$,  was derived by 
\cite[eq. 10]{Fenton1972}.

Note that, in its present form, equation \eqref{CCetabis} is not suitable 
for accurate numerical computations of $\eta$ due to the complicated 
pseudo-differential operator. For this purpose, its integral formulation is 
better suited \cite[\S6]{Clamond2018}. 
However, equations \eqref{CCeta} and \eqref{CCetabis} are convenient to 
derive analytic approximations (c.f. section \ref{appapp} where surface 
recovery is performed analytically for linear flexural-capillary-gravity waves 
in order to illustrate the procedure).

\section{Example 1: Recovery of linear flexural-capillary-gravity waves}\label{appapp}

Here, we illustrate the recovery procedure for an infinitesimal 
flexural-capillary-gravity wave that is analytically tractable via its 
linear approximation. 

\subsection{Linear approximation of a traveling wave}

For infinitesimal waves, the surface pressure \eqref{defpscapflex} and 
the Babenko-like equation \eqref{CCeta} are linearised as 
\refstepcounter{equation}
\[
\label{CCetalin}
\sur{p} \approx D \eta_{xxxx} - \tau \eta_{xx}, \qquad \sur{p} +
g \eta \approx \partial_x \cos\!\left( \depth\partial_x \right)
\sin\!\left( \depth\partial_x \right)^{-1} B \eta. 
\eqno{(\theequation{\mathit{a},\mathit{b}})}
\]
$(2\pi/k)$-periodic solutions are thus $\eta\approx a\cos(kx-\varphi)$ 
($ka\ll1$ and $\varphi$ a constant phase shift) with the (linear) dispersion 
relation
\begin{equation}
B \approx \left(g + \tau k^2 + D k^4\right)k^{-1}\tanh(k\depth) .
\end{equation}
The linear approximation of the bottom-pressure can then be 
obtained as
\refstepcounter{equation}
\[
\bot{p} \approx gd + \mathfrak{p} \cos(kx-\varphi), \quad
\mathfrak{p} = a \left( g + \tau k^2 + D k^4 \right)
\operatorname{sech}(kd) = kaB \operatorname{csch}(k\depth),
\eqno{(\theequation{\mathit{a},\mathit{b}})}\label{linsol}
\]
and the horizontal velocity at the bottom as
\begin{equation}\label{ublinsol}
\bot{u} \approx \pm \sqrt{B} \left[ 1 - B^{-1} \mathfrak{p} 
\cos(kx-\varphi) \right]\quad\implies\quad
\left< \bot{u} \right> \approx \pm \sqrt{B}.
\end{equation}
This relation shows that, to this order of approximation, the Bernoulli 
constant $B$ can be replaced by $\left<\bot{u}\right>^2$. Moreover, the  
sign of $\left<\bot{u}\right>$ gives the direction of propagation.  
Thus, in terms of parameters measurable at the bottom,  the (linearised) 
free surface is
\begin{equation}\label{solinub}
\eta \approx k^{-1} \left< \bot{u} \right>^{-2} \mathfrak{p} 
\sinh(k\depth)\cos(kx-\varphi).
\end{equation}

\subsection{Free-surface and surface-pressure recoveries}

Suppose that data of the bottom-pressure can be well approximated by the 
ansatz (\ref{linsol}{\it a}). A least squares (for example) minimisation  
between the data and (\ref{linsol}{\it a}) gives $g\depth$, $k$, 
$\varphi$ and $\mathfrak{p}$; these parameters are now definitely known. 
We have to the first-order in $\eta$ 
\begin{align}
\sur{\widetilde{\mathfrak{P}}} &\approx \mathfrak{p} \cos(kx-\varphi
+\ui k\depth) - \ui\mathfrak{p}\sin(kx - \varphi + \ui k\depth) k \eta, \\
\sur{\mathfrak{Q}} &\approx k^{-1} \mathfrak{p} \sin(kx -
\varphi + \ui k\depth) + \ui \mathfrak{p} \cos(kx - \varphi - \ui 
k\depth) \eta,
\end{align}
and, for infinitesimal waves, both $\mathfrak{p}$ and $\eta$ are small quantities 
of the same order. Thus, to the leading order, the recovery formula \eqref{recoveq} yields
\begin{equation}\label{solineta}
\mathfrak{p} \sinh(k\depth) \sin(kx - \varphi) + B \eta_x \approx 0
\quad\implies\quad
\eta = (kB)^{-1} \mathfrak{p} \sinh(k\depth) \cos(kx-\varphi),
\end{equation}
where the resolution is performed under the condition \eqref{defmean}. 
Similarly, to the leading order, the relation \eqref{psrecov} yields the 
surface-pressure
\begin{equation}\label{solinps}
\sur{p} \approx \left[ \cosh(k\depth) - g(kB)^{-1} \sinh(k\depth) \right]
\mathfrak{p} \cos(kx-\varphi) = \left[ kB \coth(k\depth) - g\right] \eta.
\end{equation}

With \eqref{solineta} and \eqref{solinps} the free-surface and the surface-pressure, 
respectively, are recovered modulo only one yet unknown parameter: the Bernoulli 
constant $B$. If, for instance, $\left<\bot{u}\right>$ has also been measured, 
then we have $B\approx\left<\bot{u}\right>^2$ and the solution \eqref{solinub} 
is recovered.
If no extra measurements are available, but if we know that we are 
dealing with flexural-capillary-gravity waves, the relation 
\eqref{defpscapflex} should apply. Thus, the quadratic error $E$ 
between   \eqref{defpscapflex} and 
\eqref{solinps} is, to the leading order,
\begin{equation}
E \approx \half \mathfrak{p}^2 \cosh(k\depth)^2
\left[ 1 - (g + \tau k^2 + D k^4) k^{-1} B^{-1} \tanh(k\depth) \right]^2,
\end{equation}
so this error is minimum if $B = ( g + \tau k^2 + D k^4) k^{-1} 
\tanh(k\depth)$, as expected. Alternatively, from the recovered 
surface pressure ${\sur{p}}_r$ given by \eqref{solinps}, we have 
$\max({\sur{p}}_r)-\min({\sur{p}}_r)=2\left(\coth(k\depth)- 
g/kB\right)\sinh(kd)\mathfrak{p}$, while the theoretical surface-pressure 
${\sur{p}}_t$ \eqref{defpscapflex} yields $\max({\sur{p}}_t)-
\min({\sur{p}}_t)\approx2(\tau k^2 + D k^4)\sinh(kd)\mathfrak{p}/kB$.
Equating these two quantities gives the expected dispersion relation.

\section{Example 2: Recovery of nonlinear capillary-gravity waves}\label{secex2}


\begin{figure}[t!]
\centering
\sidesubfloat[]{
\includegraphics*[width=.9\textwidth]{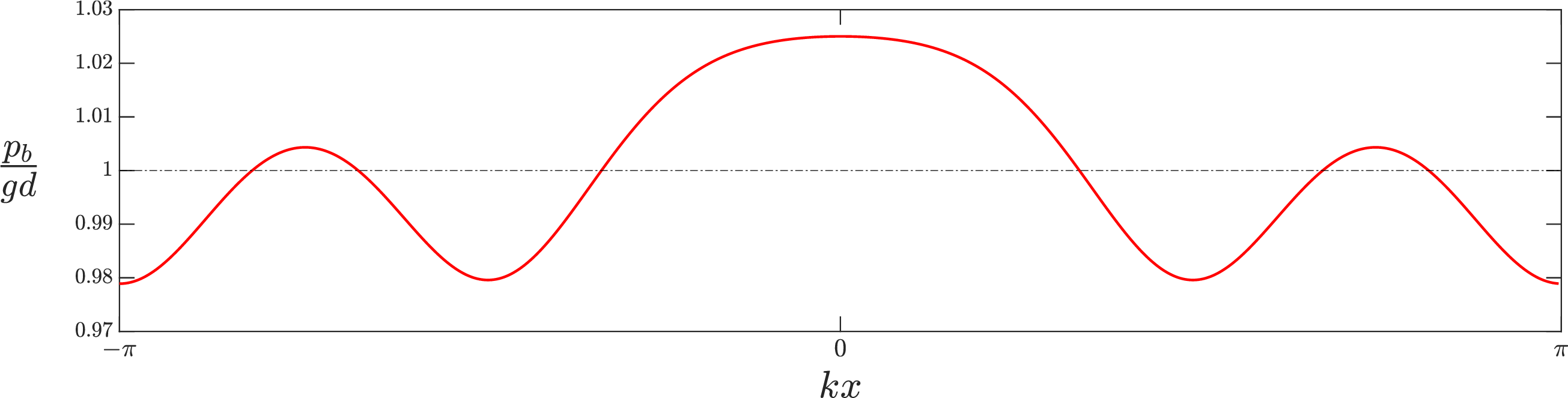}\label{Fig1a}
} \\
\sidesubfloat[]{
\includegraphics*[width=.9\textwidth]{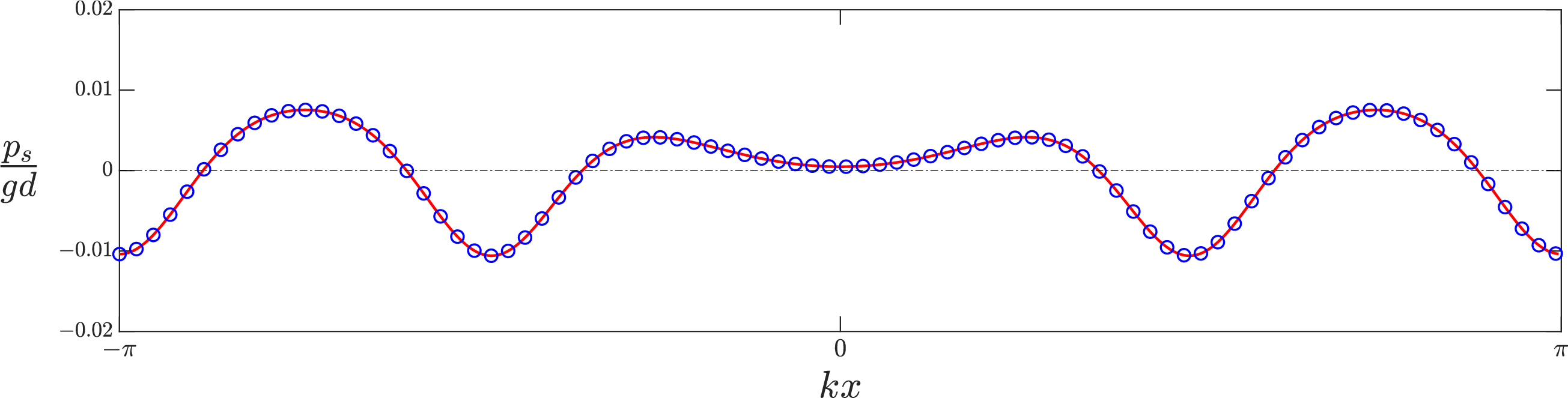}\label{Fig1b}
} \\ \hspace{.05em}
\sidesubfloat[]{
\includegraphics*[width=.895\textwidth]{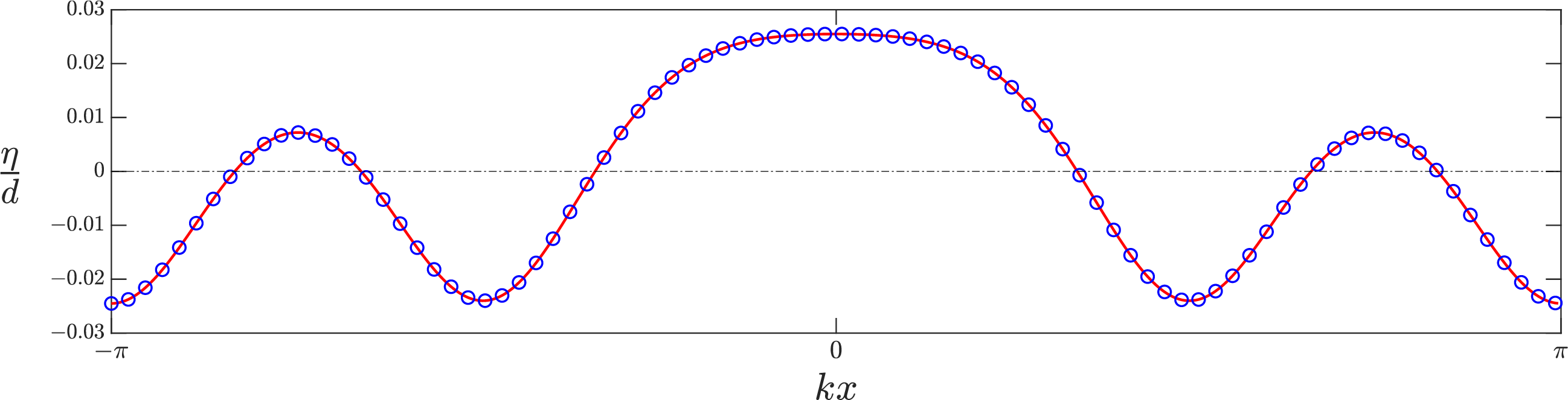}\label{Fig1c}
}

\caption{Recovery of a nonlinear capillary-gravity 
wave with period $L/\depth=6\pi$, Froude number 
square $B/gd=1.01568$ and Bond number $\tau/gd^2=1/3$. 
(a): Bottom pressure treated as a ``measurement'' 
for the recovery procedure. (b,c): Respectively, recovered 
surface pressure and profile (blue circles) versus the 
exact solution  (red line).}
\label{Fig1}
\end{figure}


\begin{figure}[t!]
\centering
\sidesubfloat[]{
\includegraphics*[width=.9\textwidth]{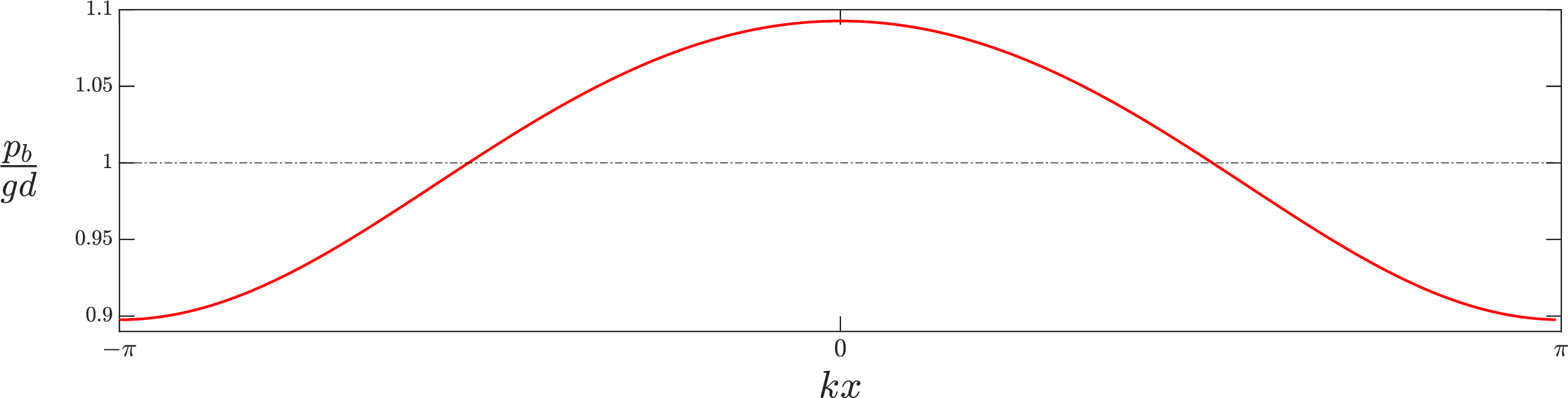}\label{Fig2a}
} \\
\sidesubfloat[]{
\includegraphics*[width=.9\textwidth]{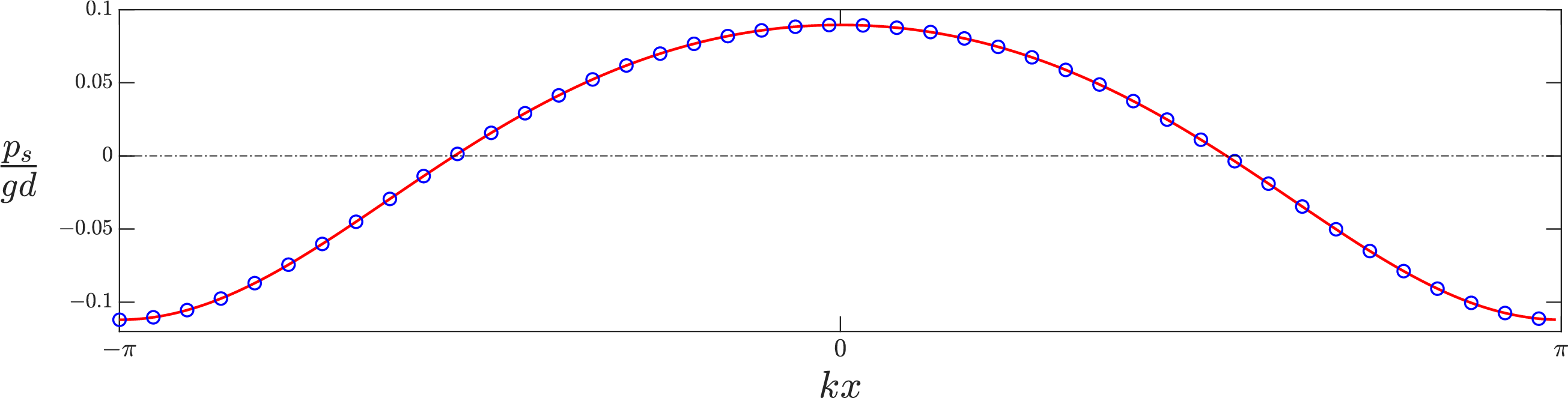}\label{Fig2b}
} \\ \hspace{.05em}
\sidesubfloat[]{
\includegraphics*[width=.895\textwidth]{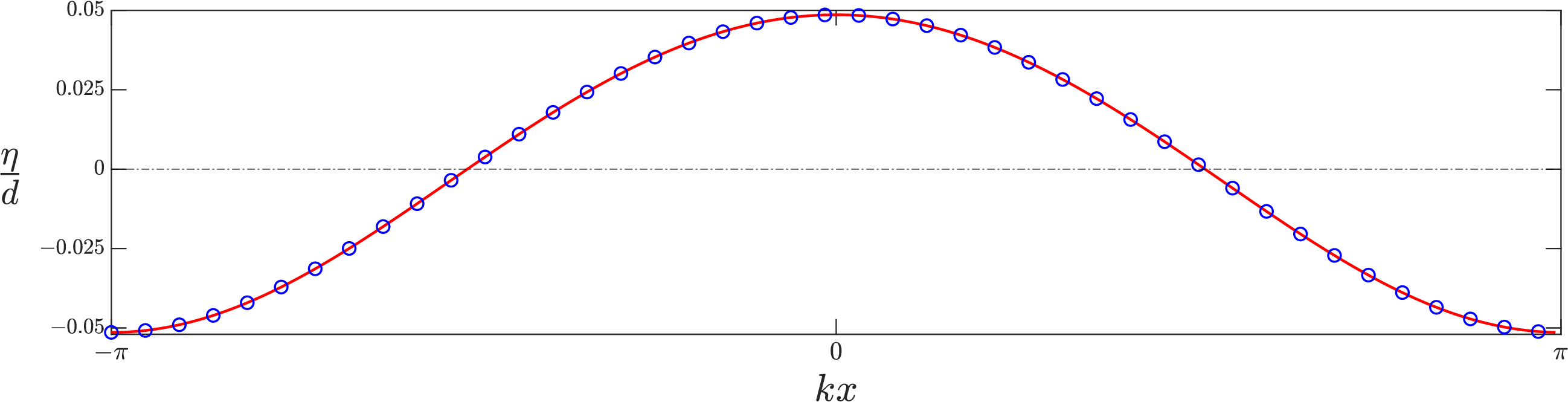}\label{Fig2c}
}

\caption{Same panels as Figure \ref{Fig1} for the 
period $L/\depth=2\pi$, the Froude number square  
$B/g\depth=2.28113$ and the Bond number $\tau/gd^2=2$.}
\label{Fig2}
\end{figure}


We now consider the fully nonlinear recovery problem for 
capillary-gravity waves. Since we do not have experimental 
data for this problem, we first compute a travelling wave 
from which we extract the bottom pressure numerically. The
algorithm used for such a computation is an adaptation of 
the method described in \cite{Clamond2018,LC23} 
when arbitrary pressure is present at the free surface.
Once computed, this accurate numerical solution is taken 
as data for the bottom pressure to reconstruct the wave 
profile, the surface pressure and various hydrodynamic 
parameters.

Following \cite{Clamond2013}, we start by expanding the 
pressure data in truncated Fourier series (collocated at 
a set of equispaced points) and perform analytic continuation 
in the complex plane \cite{Clamond2013}
\begin{equation}
\widetilde{\mathfrak{P}}(z) = \bot{p}(z+\ui\depth) - g\depth 
\approx \sum_{|n|>0}^N \mathfrak{p}_n \ue^{\ui n k (z + \ui d)} 
= \sum_{|n|>0}^N \mathfrak{p}_n \ue^{- n k \depth} \ue^{\ui n k z} .
\end{equation}
From the above definition, we compute the anti-derivative at 
the surface
\begin{equation}
\sur{\mathfrak{Q}}(x) = \int_{0}^{x} \sur{\widetilde{\mathfrak{P}}}(x') \ud x' \approx \sum_{|n|>0}^N \frac{\ui \mathfrak{p}_n}
{n k} \frac{\ue^{-nka} - \ue^{\ui nk(x+\ui\eta)}}{\ue^{nkd}} .
\end{equation}

We note that $N=256$ is sufficient enough to accurately resolve 
the Fourier spectrum (up to computer precision) of the bottom 
pressure data. Once the holomorphic functions are computed, we 
solve expression \eqref{recoveqin} by imposing the total height 
of the wave as a closure relation within the built-in iterative 
solver \textsf{fsolve} from \textsc{Matlab} \cite{ClamondEtAl2023}. 
As an initial guess, we use the linear approximation given by \eqref{solineta}. 
The algorithm only takes few seconds to run on a classical desktop 
and achieve a tolerance criterion of $\epsilon<10^{-12}$ on the residual.

We present in Figures \ref{Fig1} and \ref{Fig2} two examples of 
nonlinear capillary-gravity waves. The primary possesses a surface 
tension coefficient with critical Bond number $\textrm{Bo}\eqdef 
\tau/(gd^2)=1/3$, whereas the second is subject to strong capillary 
effects with $\textrm{Bo}=2$.
The first configuration displayed in Figure \ref{Fig1} is in rather 
shallow water, with Froude number squared $\textrm{Fr}^2\eqdef B/g
\depth=1.01568$. As clearly demonstrated in panels \ref{Fig1a} and 
\ref{Fig1b}, both highlight excellent agreement between the recovered 
surface pressure and wave profile with the solutions of reference. 
For the first case, the numerical error between recovered (r) and 
theoretically predicted (t) fields are as follows:
$\vert\vert \eta_r - \eta_t 
\vert\vert_{\infty} = 6.5289\times10^{-9}$,
$\vert\vert {\sur{p}}_r - {\sur{p}}_t \vert\vert_{\infty} = 
2.8887\times10^{-8} $ and $\vert B_r - B_t 
\vert = 2.5233\times10^{-7}$.
Similarly, the second case represents waves over a significantly deep layer, where 
the inverse problem is essentially more difficult to solve as it is mathematically ill-posed. 
Nevertheless, it also shows remarkable agreement in the 
recovered data. Regarding numerical errors for this case, it yields 
$\vert\vert \eta_r - \eta_t 
\vert\vert_{\infty} = 1.3882\times10^{-9}$,
$\vert\vert {\sur{p}}_r - {\sur{p}}_t \vert\vert_{\infty} = 
1.8098\times10^{-8} $ and $\vert B_r - B_t 
\vert = 2.1645\times10^{-7}$.
We note that the Froude number square is $B/g\depth=2.28113$ in this case.

These recoveries were obtained assuming no {\em a priori} knowledge 
of the physical nature of the surface pressure, but assuming that the 
total wave height has been measured in addition to the bottom pressure. 
If instead of the total wave height we consider, say, the mean horizontal 
velocity at the bottom, we were also able to recover both the free-surface 
and surface-pressure, with similar accuracy for $\eta$ ($\sim10^{-8}$) 
and $B$ ($\sim10^{-10}$). 

With knowledge of the physical nature of the surface pressure, 
we were also able to recover the free surface without extra 
measurements beside the bottom pressure. This is obtained 
minimising an error between the reconstructed and theoretical 
surface pressure as explained in section \ref{secclosure}. Our 
preliminary numerical investigations seem to indicate that the 
choice of the error to minimise plays a role in the speed and 
accuracy of the recovery procedure. A thorough numerical 
investigation of this optimisation problem is way beyond the 
scope of this short paper, which purpose is a proof of concept 
to attest the possibility to recover both the free-surface and 
the surface-pressure.


\section{Discussion}\label{secconclu}

We derived expressions for free-surface and surface-pressure recoveries, 
assuming the physical effects at the free surface or considering additional 
measurements.
Then, we illustrated the practical procedure with a fast and simple numerical 
algorithm. The method proposed here is more general in substance than previous
studies 
by \cite{ClamondConstantin2013,Clamond2013,ClamondHenry2020}, and can be 
generalised to incorporate linear shear currents along the lines 
of \cite{ClamondEtAl2023}. This approach can further be extended 
to accommodate overhanging waves (existing in presence of capillary and/or 
vorticity) as recently shown by \cite{LC23}. 

So far, we have considered recovery procedures from bottom pressure measurements, 
but similar relations could be derived considering the pressure at another depth, 
as well as other measured physical quantities.   
Further extensions to configurations with non-permanent 
wave motions or arbitrary vorticity, for example, are also of great interest, 
but present technical challenges beyond the scope of this current work. 

In this short paper, we demonstrated the possibility to recover the free-surface 
with arbitrary surface-pressure, and we briefly illustrated the procedure with 
few examples. We did not address the (difficult) question of uniqueness of 
the free-surface from a given bottom-pressure. Indeed, for instance,  
capillary-gravity waves are not unique for identical physical parameters 
\cite{BuffoniEtAl1996,ClamondEtAl2015}. This example indicates, although the 
recovery from bottom measurements is a slightly different problem, that the 
question of uniqueness is important, both theoretically and practically, and 
it should be the subject of future investigations. \\


\noindent{\bf Funding.} Joris Labarbe has been supported by the French government, 
through the $\mbox{UCA}^{\mbox{\tiny JEDI}}$ {\em Investments in the Future\/} 
project managed by the National Research Agency (ANR) with the reference number 
ANR-15-IDEX-01. \\

\noindent{\bf Declaration of interests.} The authors report no conflict of interest.

\bibliographystyle{amsplain}
\bibliography{Biblio}

\providecommand{\bysame}{\leavevmode\hbox to3em{\hrulefill}\thinspace}
\providecommand{\MR}{\relax\ifhmode\unskip\space\fi MR }
\providecommand{\MRhref}[2]{%
  \href{http://www.ams.org/mathscinet-getitem?mr=#1}{#2}
}
\providecommand{\href}[2]{#2}
\begin{thebibliography}{10}

\bibitem{Babenko1987}
K.~I. Babenko, \emph{{Some remarks on the theory of surface waves of finite amplitude}}, Sov. Math. Dokl. \textbf{35} (1987), 599--603.

\bibitem{BuffoniEtAl1996}
B.~Buffoni, M.~D. Groves, and J.~F. Toland, \emph{A plethora of solitary gravity-capillary water waves with nearly critical {B}ond and {F}roude numbers}, Phil. Trans. R. Soc. Lond. A \textbf{354} (1996), no.~1707, 575--607.

\bibitem{Clamond2013}
D.~Clamond, \emph{{New exact relations for easy recovery of steady wave profiles from bottom pressure measurements}}, J. Fluid Mech. \textbf{726} (2013), 547--558.

\bibitem{Clamond2018}
\bysame, \emph{New exact relations for steady irrotational two-dimensional gravity and capillary surface waves}, Phil. Trans. R. Soc. A \textbf{376} (2018), no.~2111, 20170220.

\bibitem{Clamond2022}
\bysame, \emph{Explicit {D}irichlet--{N}eumann operator for water waves}, J. Fluid Mech. \textbf{950} (2022), A33.

\bibitem{ClamondConstantin2013}
D.~Clamond and A.~Constantin, \emph{Recovery of steady periodic wave profiles from pressure measurements at the bed}, J. Fluid Mech. \textbf{714} (2013), 463--475.

\bibitem{ClamondEtAl2015}
D.~Clamond, D.~Dutykh, and A.~Dur{\'{a}}n, \emph{{A plethora of generalised solitary gravity-capillary water waves}}, J. Fluid Mech. \textbf{784} (2015), 664--680.

\bibitem{ClamondHenry2020}
D.~Clamond and D.~Henry, \emph{Extreme water wave profile recovery from pressure measurements at the seabed}, J. Fluid Mech. \textbf{903} (2020), R3.

\bibitem{ClamondEtAl2023}
D.~Clamond, J.~Labarbe, and D.~Henry, \emph{Recovery of steady rotational wave profiles from pressure measurements at the bed}, J. Fluid Mech. \textbf{961} (2023), R2.

\bibitem{Constantin2012b}
A.~Constantin, \emph{On the recovery of solitary wave profiles from pressure measurements.}, J. Fluid Mech. \textbf{699} (2012), 376--384.

\bibitem{Fenton1972}
J.~D. Fenton, \emph{A ninth-order solution for the solitary wave}, J. Fluid Mech. \textbf{53} (1972), no.~2, 257--271.

\bibitem{LC23}
J.~Labarbe and D.~Clamond, \emph{General procedure for free-surface recovery from bottom pressure measurements: {A}pplication to rotational overhanging waves}, J. Fluid Mech. (2023), no.~in press. arXiv:2308.10567.

\bibitem{Lagrange1781}
J.-L. Lagrange, \emph{M\'emoire sur la th\'eorie du mouvement des fluides}, Nouv. M{\'{e}}m. Acad. Berlin (1781), 151--198.

\bibitem{LAmb1932}
H.~Lamb, \emph{Hydrodynamics}, 6th ed., Dover, 1932.

\bibitem{OliverasEtAl2012}
K.~L. Oliveras, V.~Vasan, B.~Deconinck, and D.~Henderson, \emph{Recovering surface elevation from pressure data}, SIAM J. Appl. Math. (2012), 897--918.

\bibitem{Toland2008}
J.~F. Toland, \emph{Steady periodic hydroelastic waves}, Arch. Rat. Mech. Anal. \textbf{189} (2008), 325--362.

\bibitem{VasanEtAl2017}
V.~Vasan, K.~Oliveras, D.~Henderson, and B.~Deconinck, \emph{{A method to recover water-wave profiles from pressure measurements}}, Wave Motion \textbf{75} (2017), 25--35.

\bibitem{WadeEtAl2014}
S.~L. Wade, B.~J. Binder, T.~W. Mattner, and J.~P. Denier, \emph{On the free-surface flow of very steep forced solitary waves}, J. Fluid Mech. \textbf{739} (2014), 1--21.

\bibitem{WehausenLaitone1960}
J.~V. Wehausen and E.~V. Laitone, \emph{Surface waves}, Fluid {D}ynamics {III} (S.~Flugge and C.~Truesdell, eds.), Encyclopaedia of {P}hysics, vol.~IX, Springer-Verlag, 1960, pp.~446--778.

\end{thebibliography}

\end{document}